\documentclass[prd,twocolumn,amsmath,amssymb,floatfix,superscriptaddress]{revtex4}
\usepackage{bm}
\usepackage{amsmath}
\usepackage{epsfig}
\usepackage{color}
\usepackage{natbib}
\usepackage{graphicx}
\usepackage{hyperref}
\usepackage{ifthen}
\usepackage{xstring}
\usepackage{graphicx}

\def\simlt{\lesssim}

\newcommand{\Mpch}{\mbox{Mpc}/h}

\newcommand{\refsec}[1]{section~\ref{sec:#1}}
\newcommand{\refeq}[1]{Eq.~(\ref{eqn:#1})}

\newcommand{\reffig}[1]{Fig.~\ref{fig:#1}}

\definecolor{darkgreen}{cmyk}{0.85,0.2,1.00,0.2} 
\definecolor{purple}{cmyk}{0.5,1.0,0,0}

\begin{document}
\title{Cluster Abundance in \boldmath$f(R)$ Gravity Models}

\author{Simone Ferraro}
\affiliation{Kavli Institute for Cosmological Physics, University of Chicago, Chicago IL 60637}
\author{Fabian Schmidt}
\affiliation{Theoretical Astrophysics, California Institute of Technology, Mail Code 350-17, Pasadena, CA  91125}
\author{Wayne Hu}
\affiliation{Kavli Institute for Cosmological Physics, University of Chicago, Chicago IL 60637}

\date{\today}

\begin{abstract}
\baselineskip 11pt
 As one of the most powerful probes
of cosmological structure formation, the abundance of massive galaxy clusters is a sensitive probe of modifications to gravity on cosmological scales.  In this paper, we
present results from $N$-body simulations of a general class of $f(R)$ models,
which self-consistently solve the non-linear field equation for the enhanced
forces.  Within this class we vary the amplitude of the 
field, which controls the range of the enhanced gravitational forces, both 
at the present epoch and as a function of redshift. 
Most models in the literature can be mapped onto the parameter space of this
class.  Focusing on the abundance of massive dark matter halos,   
we compare the simulation results to a simple spherical collapse model.  
Current constraints lie in the large-field regime, where the chameleon 
mechanism is not important.  In this regime, the spherical
collapse model works equally well for a wide range of models and
can serve as a model-independent tool for placing constraints on 
$f(R)$ gravity from cluster abundance.  Using these results, we show how
 constraints from the observed local abundance of X-ray clusters
on a specific $f(R)$ model can be mapped onto other members of this general class of models.  
\end{abstract}

\maketitle

\section{Introduction} \label{sec:intro}

The abundance of massive galaxy clusters provides a unique test of gravity on 
cosmological scales \cite{Schmidt:2009am,RapettiEtal,Lombriser:2010mp}.  
Once constrained to expansion history data, modified gravity explanations of the cosmic acceleration generically  predict very different effects on the
growth of cosmological structure than spatially smooth dark energy like the 
cosmological constant.  Moreover as highly non-linear objects, clusters provide
a testing ground for the non-linear interactions of viable theories where gravity becomes indistinguishable from General Relativity locally.

In the so-called $f(R)$ class of models (see \cite{Nojiri:2006ri,Sotiriou:2008rp} and references therein) the modification to gravity arises from replacing 
the Einstein-Hilbert action by a function of
the Ricci or curvature scalar $R$  \cite{Capozziello:2002rd,NojOdi03,Capozziello:2003tk}.
These models possess an extra scalar degree of freedom $f_R \equiv df/dR$ which
mediates a 4/3 enhancement of gravitational forces on scales below the Compton 
wavelength or range associated with its mass.    

This enhancement changes the abundance
of rare dark matter halos associated with clusters of galaxies.   Measurements of the
cluster abundance provide the current best cosmological constraints on $f(R)$ models
\cite{Schmidt:2009am,Lombriser:2010mp}.
On the other hand,
in order to hide these enhancements from local tests of gravity, viable $f(R)$ models
employ the chameleon mechanism which allows the Compton wavelength
to shrink 
in regions with deep gravitational potential wells \cite{Mota:2003tc,khoury04a}.  Cosmological
simulations including the chameleon effect are required to explore the impact of
these modified forces on the cluster abundance.   These have so far been performed for only a specific form of $f(R)$
 \cite{oyaizu08b,Pkpaper,halopaper}.

In fact, the relationship between the Compton wavelength, chameleon threshold and their
respective evolution with redshift depends on the functional form of $f(R)$.  
In this paper, we explore the dependence of the cluster abundance on the functional
form of $f(R)$ in order to place more robust constraints on the whole class of models.

In \S \ref{sec:meth}, we review the phenomenology of $f(R)$ models, simulation technique and spherical collapse modeling as well as show that
a general class of broken power law models introduced in Ref.~\cite{HuSaw07a} 
covers most cases of cosmological interest.  In \S \ref{sec:abundance} we
study the enhancement of the cluster abundance in these models and 
obtain constraints from the local X-ray sample.  We discuss these
results in \S \ref{sec:discussion}.

\section{Methodology}
 
We begin in \S \ref{sec:models} with a review of $f(R)$ models.  In \S \ref{sec:sims} we discuss the numerical $N$-body simulations from which we extract the 
cluster abundance enhancements.  In \S \ref{sec:sims}, we discuss the semi-analytic
modeling of these results with spherical collapse collapse calculations.

 \label{sec:meth}

\subsection{Models}
\label{sec:models}

In the  $f(R)$ model, the Einstein-Hilbert action is augmented with a general function of the scalar curvature $R$,
 \begin{eqnarray}
S_{G}  =  \int{d^4 x \sqrt{-g} \left[ \frac{R+f(R)}{16\pi G}\right]}\,. 
\label{eqn:action}
\end{eqnarray}
Here and throughout $c=\hbar=1$.    
Gravitational force enhancements are associated
with an additional scalar degree of freedom, the chameleon field $f_{R}\equiv df/dR$, and have a range
given by the comoving Compton wavelength 
$\lambda_C= a^{-1}(3 d f_R/dR)^{1/2}$. This additional attractive force
leads to the enhancement in the abundance of rare massive dark matter halos described below.    The second important property of such models is the non-linear chameleon effect which
shuts down the enhanced forces in regions with deep gravitational potential wells compared
with the field at the background curvature $\bar R$,
$|\Psi| > | f_R(\bar R)|$.

Given that different models for $f(R)$ produce different scalings of the
Compton wavelength and chameleon threshold with curvature and hence implicitly with redshift and the degree
of non-linearity, we wish to explore the dependence of the halo abundance with 
variations in the form of $f(R)$.

\begin{figure}[t]
\centerline{\psfig{file=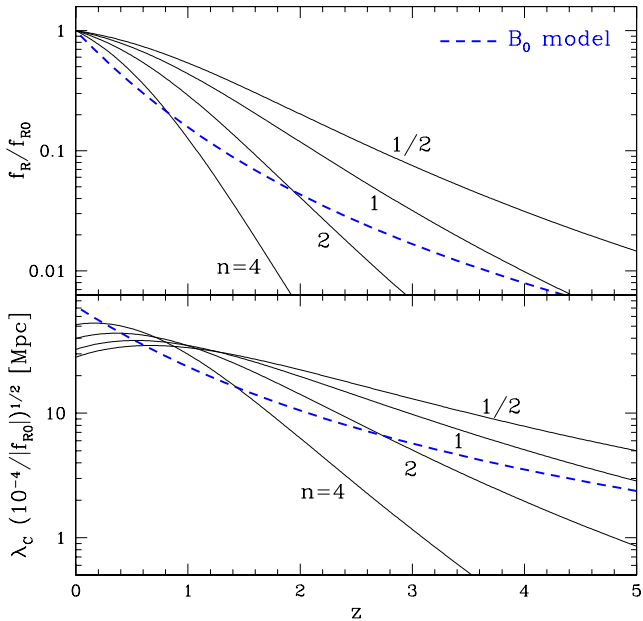, width=3.5in}}
\caption{\footnotesize Redshift evolution of the chameleon field $f_R$ 
(top panel) and Compton wavelength $\lambda_C$ (bottom panel) 
in the background for the broken power law class of models.  As the
scaling index $n$ increases the field amplitude becomes increasingly suppressed leading to stronger
chameleon effects for the same gravitational potentials of clusters.  The Compton wavelength for a fixed field amplitude today remains relatively constant at 
$z \simlt 1$ and then also becomes increasingly suppressed with $n$.  An alternative
class of models specified by the expansion history and Compton wavelength $B_0$ parameter is also shown for comparison (dashed lines).}
\label{fig:fieldwavelength}
\end{figure}

We therefore choose a class of models where the scaling index with curvature
can vary as a broken power law \cite{HuSaw07a}  such that
\begin{equation}
f(R) = - 2\Lambda \frac{R^n}{R^n+\mu^{2n}},
\end{equation}
with two free parameters, $\Lambda$, $\mu^2$ for each value of the scaling index $n$. 
Note that as $R\rightarrow 0$,
$f(R)\rightarrow 0$, and hence these models do not contain a cosmological
constant. 
Nonetheless as $R \gg \mu^2$,  the function $f(R)$ can be approximated as
\begin{eqnarray}
f(R) = -2 \Lambda -\frac{ f_{R0}}{n} \frac{\bar R_0^{n+1}}{ R^n} \,,
\label{eqn:fRapprox}
\end{eqnarray}
with $f_{R0}= -2n \Lambda \mu^{2n}/\bar R_0^{n+1}$ replacing $\mu$ as the second parameter
of the model.   
  Here we define $\bar R_{0}=\bar R(z=0)$, 
so that $f_{R0}= f_{R}(\bar R_{0})$, where overbars denote the quantities of the
background spacetime.  Note that if $|f_{R0}| \ll 1$ the curvature scales set by $\Lambda
={\cal O}(R_0)$
and $\mu^2$ differ widely and hence the $R \gg \mu^2$ approximation is valid today and for
all times in the past.

The background expansion history mimics $\Lambda$CDM with $\Lambda$ as
a true cosmological constant
to order $f_{R0}$.   Therefore in the limit 
$|f_{R0}| \ll 10^{-2}$, the $f(R)$ model and $\Lambda$CDM are essentially indistinguishable with geometric tests.   On the other hand, the field amplitude parameter 
($f_{R0}$) controls the range of the force modification and the chameleon mechanism. 
With the functional form of Eq.~(\ref{eqn:fRapprox}), the comoving Compton wavelength becomes
\begin{equation}
{\lambda_C }= a^{-1} \sqrt{ 3(n+1) | f_{R0}| {R_0^{n+1} \over R^{n+2}}} \,,
\end{equation}
with a value at the background curvature today $R_0=3H_0^2(4-3\Omega_{\rm m})$ of 
\begin{equation}
\lambda_{C0} \approx 16.6 \sqrt{{ | f_{R0}| \over 10^{-4}}
{n+1 \over 4-3\Omega_{\rm m}}} h^{-1} {\rm Mpc} \,,
\end{equation}
assuming a flat universe.
As the scaling index $n$ increases, the Compton wavelength today increases given the same
background field amplitude today $f_{R0}$.  Conversely as $n$ increases, force modifications
at high redshift versus today decrease and the chameleon mechanism extends to
shallower potential wells.  Thus the net effect is a fairly weak dependence of $\lambda_C$
 on $n$ at $z \simlt 1$.  In  Fig.~\ref{fig:fieldwavelength}, we show the evolution of
 the background field and Compton wavelength for the $\Omega_{\rm m}=1-\Omega_\Lambda=0.24$,
 $h=0.73$ cosmology that we simulate below.

This set of broken power law models covers the cosmological 
phenomenology of most viable $f(R)$ models.  For example the models of Ref.
\cite{Starobinsky:2007hu} compose a subset of this class.
It also has sufficient flexibility to bracket the behavior of 
 models where the
combination of a specific expansion history \cite{Multamaki:2005zs,Capozziello:2006dj} 
and the Compton wavelength today
fixes the form of $f(R)$ \cite{SonHuSaw07}.  For the $\Lambda$CDM expansion history and
a dimensionless Compton wavelength parameter 
\begin{equation}
B_0 \equiv {df_{R}/dR \over 1+f_R}{ R'}{H \over H'} \Big|_{z=0} \approx 2.1 \Omega_{\rm m}^{-0.76} |f_{R0}|\,,
\end{equation}
where $' \equiv d/d\ln a$,
the redshift evolution goes from $n \sim 4$ at low curvature and 
redshift to $n \sim 0.13$ at high curvature and redshift (see Fig.~\ref{fig:fieldwavelength}).   For a fixed $|f_{R0}|$ the amount of linear growth at $z=0$ in the $B_0$ model is smaller
than in the $n=1$ model and this must be borne in mind when comparing constraints
between the two models (cf.\ \cite{Schmidt:2009am,Lombriser:2010mp,Appleby:2010dx}).

Likewise  these models have stronger chameleon effects at $z\simlt 1$ than the  $n=1$ broken power law model.   A similar caveat applies to models with exponential
rather than power law suppression of the field with curvature (e.g. \cite{Appleby:2007vb}).

\subsection{Simulations}
\label{sec:sims}

We conduct $N$-body simulations of these broken power law models with a particle-mesh relaxation code
\cite{oyaizu08b,Pkpaper}.    
Briefly, at each time step we first solve the non-linear field equation for the
field fluctuation,
\begin{eqnarray}
\nabla^2 \delta  f_{R} = \frac{a^{2}}{3}\left[\delta R(f_{R}) - 8 \pi G \delta \rho_{\rm m}\right] \,,\label{eqn:frorig}
\end{eqnarray}
using a multigrid relaxation scheme.  Here coordinates are comoving, 
$\delta f_R = f_R(R)-f_R(\bar R)$,
$\delta R = R -\bar R$, $\delta \rho_{\rm m} = \rho_{\rm m} - \bar \rho_{\rm m}$.  
The $\delta f_R$ field fluctuation then acts
as an additional source to the gravitational potential,
\begin{eqnarray}
\nabla^2 \Psi = {4\pi G}a^{2} \delta \rho_{\rm m} - \frac{1}{2} \nabla^2 \delta  f_{R} \,.\label{eqn:potorig}
\end{eqnarray}
This linear equation for $\Psi$ is solved via a fast Fourier transform.  
Once $\Psi$ is known on the mesh, particles are moved in the usual way.  

Since the field equation implies that spatial variations in $\delta f_R$ will be of order
the gravitational potential, there are two regimes of interest.  In the large-field 
regime, the background value $f_R(\bar R)$ is large compared with the gravitational
potentials of structure, and the field equation~(\ref{eqn:frorig}) can be linearized via
\begin{equation}
\delta R \approx { {d R \over d f_{R}} \bigg|_{\bar R(a)} \delta f_R } = 3 \lambda_C^{-2}(a) \delta f_R \,,
\end{equation}
where $\lambda_C(a)$ is evaluated at the background curvature $\bar R(a)$.
In this case the joint solution of the Poisson and field equations in Fourier space is
\begin{equation}
k^2 \Psi = -4 \pi G \left( {4 \over 3} - {1 \over 3} {1 \over k^2\lambda_C^2(a) + 1} \right) a^2  \delta \rho_{\rm m}. 
\label{eqn:nochameleon}
\end{equation}
Hence the background Compton wavelength sets the global range of the enhanced
gravitational force.
We call this the {\em{no chameleon}} case and for comparison conduct separate simulations
employing  Eq.~(\ref{eqn:nochameleon}).

In the small-field regime, $f_R(\bar R)$ is comparable to or smaller than typical gravitational
potentials of structure, so that the curvature changes non-linearly with the field. In other
words the  Compton wavelength depends on the local curvature or field 
$\lambda_C = \lambda_C(a,{\bf x})$.  
Field fluctuations saturate in deep gravitational potential wells ($f_R \rightarrow 0$), leading to 
an equilibrium solution $\delta R = 8\pi G   \delta \rho_{\rm m}$ and a suppression of
non-Newtonian forces.

We use simulations of three different box sizes ($400, 256, 128 \Mpch$),
and 6 simulations for each box size and model.  The runs and models
as well as mass resolution for each box are summarized in Table~\ref{table:runs}.  
To reduce the effect of sample variance, we compare each $f(R)$ simulation run
to a $\Lambda$CDM simulation with the same initial conditions, 
i.e. the same initial density field drawn from an initial power 
spectrum with $A_s = (4.73\times 10^{-5})^2$ at $k=0.05$Mpc$^{-1}$ and
$n_s=0.958$.

We measure the mass function from the simulations using the methodology described in
\cite{halopaper} and refer the reader to details therein.  Briefly, we identify halos
using a spherical overdensity criterion of $\Delta = 200$ with respect to the mean density
 and quantify the mass function enhancements
of the $f(R)$ models over $\Lambda$CDM with the same initial conditions.  
To reduce the effect of shot noise we bin results into coarse mass intervals
corresponding to approximately an e-fold ($\Delta\ln M_{200}=1.04$).
Furthermore, due to resolution effects, we only utilize halos that contain
at least 800 particles corresponding to the minimum mass given in Tab.~\ref{table:runs}.

We estimate sampling errors via bootstrap resampling.  
Note that due to our limited number of realizations, these errors might
be underestimated at high masses where halos are rare and fluctuations
are significant.

\begin{table}
\caption{Summary of simulations used for this work} 
\begin{center}
  \leavevmode
  \begin{tabular}{c|c|c c c}
  && \multicolumn{3}{|c}{$L_{\rm box}$ ($h^{-1}$ Mpc)} \\ 
  
&$|f_{R0}|$ &\ \ $400$\ \ & $256 $\ \ \  & $128$\ \ \   \\
\hline
\# full\ \ & $10^{-4}$ ($n$=1, 2) & 6 & 6 & 6 \\
runs \ & $3 \cdot 10^{-6}$ ($n$=2) & 6 & 6 & 6 \\
\ \ & $10^{-6}$ ($n$=1)  & 6 & 6 & 6 \\
\hline
\# no \ \ & $10^{-4}$ ($n$=1/2, 1, 2, 4, 8) & 6 & 6 & 6 \\
cham.~runs \ & $3 \cdot 10^{-6}$ ($n$=2)  & 6 & 6 & 6 \\
\ \ & $10^{-6}$ ($n$=1)  & 6 & 6 & 6 \\
\hline
$\Lambda$CDM &0  & 6     & 6     & 6     \\
\hline
\multicolumn{2}{c|}{$M_{\rm h, min}$ ($10^{12} h^{-1} M_\odot$)\ \ } & 204  & 53.7 & 6.61 \\
\hline
\end{tabular}
\end{center}
\label{table:runs}
\end{table}

\subsection{Spherical Collapse Predictions}
\label{sec:collapse}

Since the large field regime is where the current local cluster abundance 
measurements constrain
$f(R)$ models \cite{Schmidt:2009am,Lombriser:2010mp}, characterizing this regime in a way that does not require simulations of each model is important.   We briefly review a method utilizing spherical collapse introduced in Ref.~\cite{halopaper}

The Sheth-Tormen  description 
 for the comoving number density of halos per logarithmic interval in the virial mass $M_{\rm v}$ is given by
\begin{align}
n_{\ln M_{\rm v}} \equiv
\frac{d n}{d\ln M_{\rm v}} &= {\bar \rho_{\rm m} \over M_{\rm v}} f(\nu) {d\nu \over d\ln M_{\rm v}}\,, 
         \label{eqn:massfn}
\end{align}
where the peak threshold $\nu = \delta_c/\sigma(M_{\rm v})$ and 
\begin{eqnarray}
\nu f(\nu) = A\sqrt{{2 \over \pi} a\nu^2 } [1+(a\nu^2)^{-p}] \exp[-a\nu^2/2]\,.
\end{eqnarray}
Here
$\sigma(M)$ is the variance of the linear density field convolved with a top hat of radius $r$
that encloses $M=4\pi r^3 \bar \rho_{\rm m}/3$ at the background density
\begin{eqnarray}
\sigma^2(r) = \int \frac{d^3k}{(2\pi)^3} |\tilde{W}(kr)|^2 P_{\rm L}(k)\,,
\label{eqn:sigmaR}
\end{eqnarray}
where $P_{\rm L}(k)$ is the linear power spectrum and $\tilde W$ is the Fourier transform
of the top hat window.  The normalization constant $A$ is chosen 
such that $\int d\nu f(\nu)=1$. The parameter values of $p=0.3$, $a=0.75$, and
$\delta_c=1.673$ for the spherical collapse threshold have previously been shown to 
match simulations of $\Lambda$CDM at the $10-20\%$ level. 
The virial mass is defined as the mass enclosed at 
the virial radius $r_{\rm v}$, at which the average density is $\Delta_{\rm v}$
times the mean density.   The virial mass can then be transformed to alternate
overdensity criteria assuming a Navarro-Frenk-White density profile
\cite{HuKravtsov}.

Spherical collapse can also provide a model for the mass function enhancement
measured in the $f(R)$ $N$-body simulations \cite{halopaper}. The mass function calculation
again uses the Sheth-Tormen form of Eq.~(\ref{eqn:massfn}) but with the linear power
spectrum for the $f(R)$ model in Eq.~(\ref{eqn:sigmaR}), and two limiting cases for the spherical
collapse parameters. In one case, we simply assume that the
spherical perturbation considered is always larger than the 
Compton wavelength of the $f_R$ field, so that gravity is GR throughout, 
and the spherical
collapse parameters are unchanged. In the second case, we assume that the perturbation is
always smaller than the local Compton wavelength in spite of the redshift
evolution of the background Compton wavelength and chameleon mechanism (see
Fig.~\ref{fig:fieldwavelength}).
Hence forces are simply universally enhanced by 4/3.  In both cases, we use the modified linear force calculation for
the linear power spectrum and $\sigma(M)$ via Eq.~(\ref{eqn:massfn}).   
Hence, unmodified spherical collapse \emph{parameters} does not equate to unmodified
spherical collapse \emph{predictions}.  

The values of the resulting linear collapse threshold $\delta_c$ and virial
overdensity $\Delta_{\rm v}$ are summarized in Table~\ref{table:sc_params}.  
We use the GR values to calculate the mass function \refeq{massfn} in terms of
virial mass $M_{\rm v}$ ($M_{\rm v} \equiv M_{\Delta_{\rm v}}$)  for
$\Lambda$CDM, and correspondingly for $f(R)$ with either set of collapse
parameters.  We then rescale both
mass functions to our adopted mass definition $M_{200}$ and convolve them
with the mass binning used in the simulations before taking the ratio.  

\begin{table}[b]
\caption{Spherical collapse parameters} 
\begin{center}
  \leavevmode
  \begin{tabular}{c | c c | c c}
  & \multicolumn{2}{|c}{$z=0$} 
  & \multicolumn{2}{|c}{$z=0.316$} \\
\cline{2-5}
 & GR & mod. forces & GR & mod. forces \\
\hline
$\delta_c$  & 1.673 & 1.692 & 1.679 & 1.697 \\
$\Delta_v$  & 391 & 309 & 279 & 222 \\
\hline
\end{tabular}
\end{center}
\label{table:sc_params}
\end{table}

\begin{figure}[t]
\centerline{\psfig{file=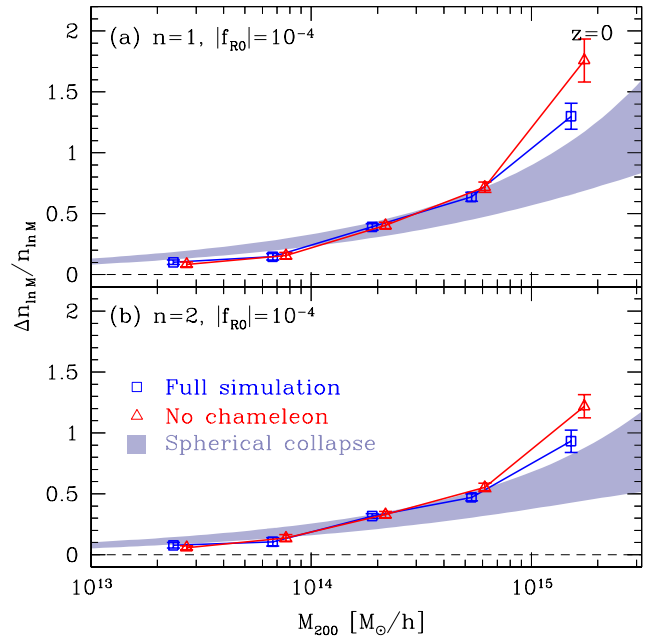, width=3.5in}}
\caption{\footnotesize Mass function enhancement at $z=0$ in large field $|f_{R0}|=10^{-4}$ models for scaling index of $n=1,2$ relative to $\Lambda$CDM.  
Here and in the following figures, the no-chameleon results have been displaced horizontally for clarity.  
Enhancement depends mainly on mass due to the increasing rarity of high mass halos.  
As $n$ increases, the enhancement drops only moderately given the small
change in the background Compton wavelength at $z \simlt 1$, consistent with only a small contribution from the non-linear chameleon effect.  
The spherical collapse predictions (shaded range) capture these qualitative trends and 
provide conservative lower limits to the enhancement.}
\label{fig:mfnlarge}
\end{figure}

\begin{figure}[t]
\centerline{\psfig{file=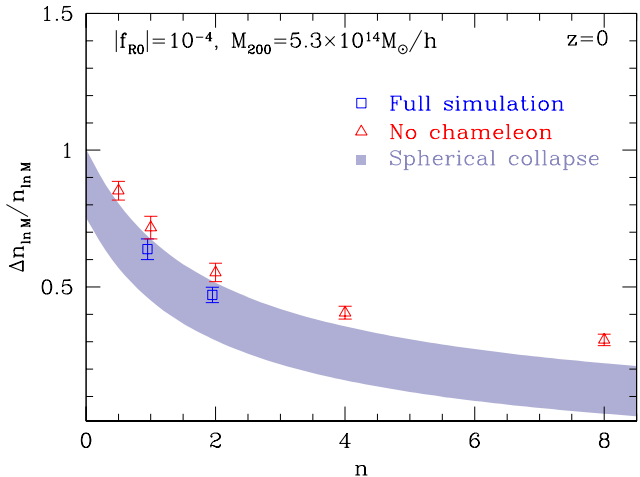, width=3.5in}}
\caption{\footnotesize Mass function enhancement at $z=0$ as a function of scaling index $n$
for the mass bin centered at $M_{200}=5.3\times 10^{14} M_\odot/h$ for large
field models $|f_{R0}|=10^{-4}$.  Spherical collapse predictions (shaded) capture the trend in the no-chameleon simulations.
Full results for $n=1,2$ and consideration of the field evolution suggests that spherical
collapse predictions should hold for $n \simlt 4$.  
Note that errors are fully correlated in that the all simulations use the same initial conditions and are compared against the same set of $\Lambda$CDM simulations.}
\label{fig:nplot}
\end{figure}

\section{Cluster Abundance} \label{sec:abundance}

With the $f(R)$ simulations described in Tab.~\ref{table:runs}, we can now test the
model dependence of the cluster abundance enhancement as well as the
accuracy of the model-independent spherical collapse technique described in the
previous section.  In \S \ref{sec:large} we discuss the large field regime relevant
for current constraints from clusters.  In \S \ref{sec:small} we evaluate the impact
of the non-linear chameleon mechanism in the small field regime.   Finally we
show how constraints on one $f(R)$ model can be transformed to another using
simulation calibrated spherical collapse methods in \S \ref{sec:current}.

\subsection{Large Field Regime}
\label{sec:large}

In Fig.~\ref{fig:mfnlarge}, we show the mass function enhancements for a large field case $|f_{R0}|=10^{-4}$ 
for $n=1,2$.  Note that we plot the data points at the center of each mass bin,
while the average mass of halos within the bin is generally smaller than that due to the steepness of the mass function.  
The spherical collapse predictions are convolved with the mass bin and hence
take into account this effect.  The uppermost mass bin extends to infinite mass
so as to include all remaining halos but is still plotted at 
$\Delta\ln M_{200}=1.04$ above the previous bin.

As the mass increases and halos become rarer in the $\Lambda$CDM simulations, 
the fractional impact of the force enhancement on
cluster abundance increases.  Relative to this overall enhancement the impact
of changing the scaling parameter $n$ is less significant.  
This weak dependence is in spite of 
 the rapid change in the background $f_{R}$ field
shown in Fig.~\ref{fig:fieldwavelength}.

We can understand this relative insensitivity by comparing the full simulation results
to the no-chameleon simulations where the
Compton wavelength is fixed to its background value through Eq.~(\ref{eqn:nochameleon}).  
Mass function enhancements in the chameleon and no-chameleon simulations are nearly the same up until the very highest masses.   For the large field value today  $|f_{R0}|=10^{-4}$, cluster potential wells are not sufficiently deep to manifest the chameleon 
mechanism today.   The small effect at the very highest masses in fact comes
from the chameleon mechanism becoming effective at high redshift as we shall see.
One can in turn understand the relative insensitivity to $n$ in the no-chameleon
simulations by examining the background Compton wavelength evolution in
Fig.~\ref{fig:fieldwavelength}.   Note that for $n \simlt 4$, the Compton wavelength
varies little for redshifts $z \simlt 1$.

\begin{figure}[t]
\centerline{\psfig{file=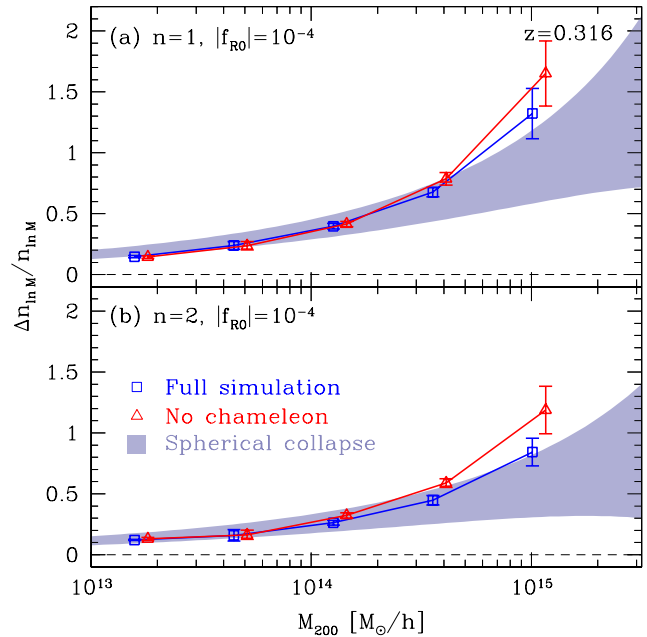, width=3.5in}}
\caption{\footnotesize Mass function enhancement at $z=0.316$ in the large field
$|f_{R0}|=10^{-4}$ for $n=1,2$. Fractional enhancements at a fixed mass remain significant at higher redshift due to the increased rarity of such halos in $\Lambda$CDM and the trend remains well captured by spherical collapse predictions (shaded region).  
Field evolution in the $n=2$ case makes the chameleon suppression in the full
simulations moderately more important.}
\label{fig:highz}
\end{figure}

The spherical collapse predictions
outlined in the previous section are also shown in Fig.~\ref{fig:mfnlarge}.  The upper boundary of the
shaded region represents enhancements predicted by the unmodified spherical
collapse parameters $\Delta_{\rm v}=391$ and $\delta_c=1.673$ whereas the lower
boundary takes the modified parameters $\Delta_{\rm v}=309$ and $\delta_c=1.692$
 (Table~\ref{table:sc_params}).
 
The spherical collapse predictions model the
results equally well for the $n=1$ and $n=2$ models.
In the high mass cluster regime, the unmodified parameters match the simulations better.  In the low
mass end the modified parameters agree better.  The modified parameters also
provide conservative estimates  of the enhancements
across the full mass range \cite{halopaper}.

We further test the large-field no-chameleon simulations against spherical collapse 
predictions for  even steeper
$n$ models in Fig.~\ref{fig:nplot}.    These predictions, based mainly on the instantaneous
linear growth function, remain accurate despite
the extremely strong scaling of the force modification with redshift in these models.  
Furthermore, Fig.~\ref{fig:fieldwavelength} implies that the no-chameleon results should
be a reasonable approximation to the full simulations for $n \lesssim 4$.
Thus, in the large field regime, one way to map cluster constraints obtained at a
given mass $M_{\rm v}$ on one $f(R)$ model to another is to match
the linear variance $\sigma(M_{\rm v})$.  
A better approximation can be obtained by setting the mass function $n_{\ln M_{\rm v}}$ equal
as we shall see in \S \ref{sec:current}.

In Fig.~\ref{fig:highz}, we show the mass function enhancements at an intermediate
redshift $z=0.316$ for the large field model.  Note that the abundance of halos of mass $M$ at $z=0$ 
is equal to that of halos of mass $M/1.5$ at this redshift in a $\Lambda$CDM
model, due to the evolution of the mass function, and we have adjusted our binning
to take this into account.   Thus for a fixed mass, the enhancement in the cluster
abundance remains significant.   Interestingly, the
 range in spherical collapse predictions continues to model these trends
 once the collapse parameters are adjusted to the matching redshift
 (see Tab.~\ref{table:sc_params}).
 
The $n=2$ results at $z=0.316$ show a slight increase in the importance
 of the chameleon suppression when compared to $z=0$ or $n=1$ at the
 same redshift.   This is consistent with the suppression of the field 
 amplitude shown in Fig.~\ref{fig:fieldwavelength}.  For $n=2$, the effect is
 only a small fractional contribution and spherical collapse predictions still work well but suggest
 that the no-chameleon approximation may have a smaller range of validity in $n$
 at high redshift.  
More generally modified gravity models which possess this type of non-linearity that 
suppresses deviations in high density regions typically do not predict larger 
enhancements of the cluster abundance at high versus low redshift at a fixed degree
of rarity or peak height $\nu$
\cite{Mortonson:2010mj}.

\begin{figure}[t]
\centerline{\psfig{file=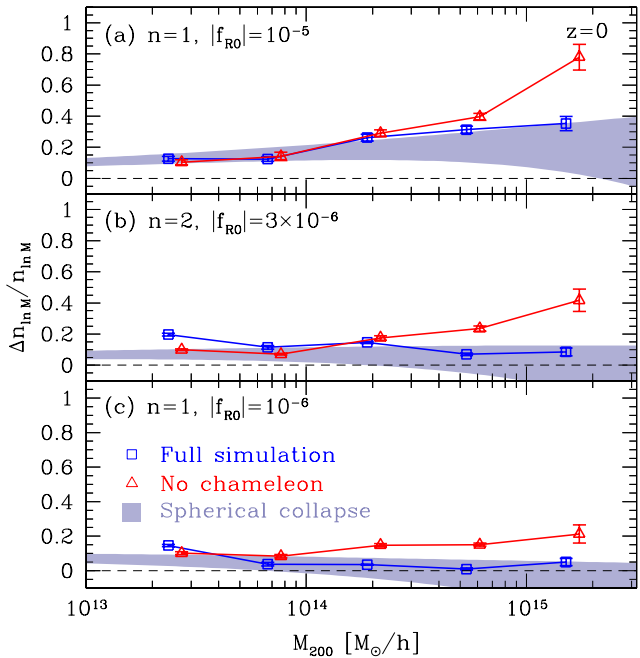, width=3.5in}}
\caption{\footnotesize Mass function enhancement in the small field regime.   The chameleon effect suppresses the enhancement when the background field amplitude
$|f_{R0}|$ drops below the depth of the gravitational potential for an object of
mass $M_{200}$.   Comparison of the full and no-chameleon simulations shows that
the limiting mass at which the chameleon appears scales roughly as expected: $M_{\rm cham} \propto |f_{R0}|^{3/2}$ nearly independently of the scaling index $n$.
Spherical collapse predictions roughly capture this suppression in the 
cluster regime $M_{200} \gtrsim 3 \times 10^{14} M_\odot/h$ but fail to
model the enhancement below $M_{\rm cham}$. }
\label{fig:mfnsmall}
\end{figure}

\subsection{Small Field Regime}
\label{sec:small}
 
As cluster abundance and other cosmological tests improve, the large-field 
models will be excluded (if no order unity excesses over $\Lambda$CDM 
expectations are detected).   In the small field regime of 
$|f_{R0}| \simlt 10^{-5}$, the chameleon mechanism is effective even today.

In Fig.~\ref{fig:mfnsmall}, we show small field results for $n=1$ and
 $|f_{R0}|=10^{-5},10^{-6}$ and, for $n=2$, $|f_{R0}|=3 \times 10^{-6}$.  
The first thing to note is that in the no chameleon simulations the impact of changing
the field value from $|f_{R0}|=10^{-4}$ to $|f_{R0}|=10^{-5}$ is less than a factor of 2
in the abundance at the highest mass bin.   We shall see in the next section, that this
logarithmic sensitivity translates into a strong model dependence of observational 
constraints on the field amplitude and Compton wavelength.

Small field results show a large
impact from the chameleon suppression as can be seen by comparing the
full simulations to the no-chameleon simulations.  
A halo is chameleon-screened whenever its gravitational potential
is larger than the field amplitude in the background $|f_{R0}|$.  This
can be used to derive a threshold mass for chameleon screening at $z=0$ for
a given value of $f_{R0}$ (see \cite{dynamicalmass}).  
We then expect the mass scale $M_{\rm cham}$ of the chameleon suppression 
in the mass function to scale similarly as the threshold for chameleon screening.  
In particular, $M_{\rm cham}$ should depend mainly on
$|f_{R0}|$ and only weakly on the scaling index $n$.  
Specifically, neglecting the small mass-dependence of the halo concentration,
we would expect the onset of the chameleon suppression to scale as 
$M_{\rm cham} \propto |f_{R0}|^{3/2}$. 

 We see from Fig.~\ref{fig:mfnsmall} that the results are consistent with this
scaling:
roughly, the chameleon for $|f_{R0}|=10^{-5}$ is important for $M_{200} \gtrsim 6 \times 10^{14} M_\odot/h$ while for $|f_{R0}|=10^{-6}$, the suppression
appears at $M_{200} \gtrsim 2 \times 10^{13} M_\odot/h$.  The $|f_{R0}|=3 \times 10^{-6}$, $n=2$ case falls consistently right in 
between the two despite being a different $n$ model.  

Spherical collapse predictions roughly model the reduced enhancement
in the cluster regime of $M_{200} \gtrsim 3 \times 10^{14} M_\odot/h$.  They correctly 
predict an absence of a significant enhancement for $|f_{R0}| \simlt 3\times 10^{-6}$. However, unmodified collapse parameter predictions can fractionally overestimate the enhancement
unlike in the large field regime, while modified collapse parameter predictions 
predict a reduction in the cluster abundance ($\Delta n_{\ln M} < 0$) not seen
in the simulations.  Moreover
both cases do not predict the correct behavior at lower masses where the full simulations
possess a higher abundance of halos than both the no-chameleon simulations 
and the collapse predictions.  Hence in the small field regime they should not be used for constraints
from galaxy groups or smaller mass objects or if precision predictions are required
at cluster masses.  We defer such modeling to 
a future work.

\begin{figure}[t]
\centerline{\psfig{file=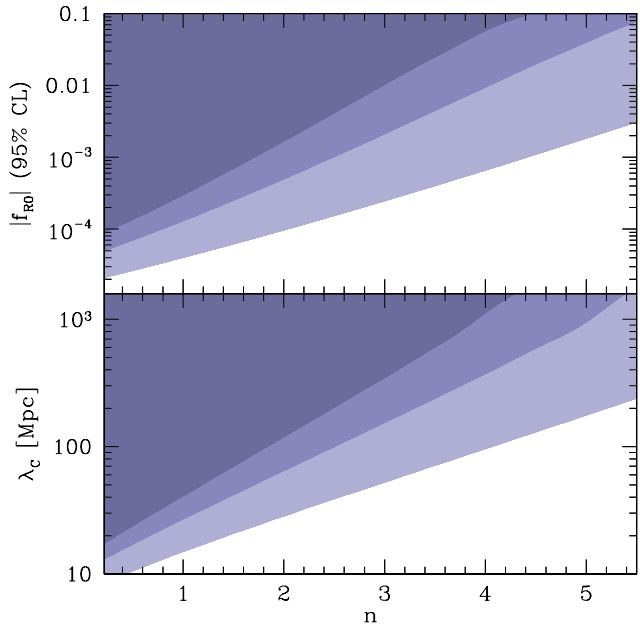, width=3.5in}}
\caption{Constraints on $|f_{R0}|$ (upper panel) and the Compton wavelength
$\lambda_C$ (lower panel) as function of the index $n$.  We have
converted the 95\% confidence level upper limits on $|f_{R0}|$ reported 
in \cite{Schmidt:2009am} for $n=1$ to other values of $n$ using the spherical
collapse model as described in the text.  The medium shaded band corresponds
to the default limit reported in \cite{Schmidt:2009am}, while dark and light
shaded areas use more or less conservative assumptions (see text). 
\label{fig:constraints}}
\end{figure}

\subsection{Current Constraints}
\label{sec:current}

Given that spherical collapse predictions work equally well for all of our broken power
law models with $n \simlt 4$ in the cluster regime and capture the scalings seen in the full 
simulations, we can extend the constraints on the 
$f(R)$ model with $n=1$ \cite{Schmidt:2009am} that were obtained using the observed abundance
of local X-ray clusters  selected in the ROSAT All-Sky Survey and followed up with Chandra
observations \cite{Vikhlinin:2008ym}.

The constraints
were obtained by using the spherical collapse model (see \refsec{collapse}) to
predict the $f(R)$ mass function enhancement at a pivot mass of
$M_{X,\rm eff} \approx 3.7\times 10^{14}\:M_{\odot}/h$, for an overdensity
of 500 with respect to critical density.  \reffig{nplot} shows that the spherical
collapse model is equally valid for other values of $n$ as long as the chameleon
effect is negligible, and it is 
straightforward to translate the constraints to other values of $n$ by
matching the abundance at $M_{X,\rm eff}$.  

The results are shown as function
of $n$ in \reffig{constraints} for a range of conservative to aggressive
interpretations of the data and modeling (see \cite{Schmidt:2009am} for further discussion).   In the top panel we show the 95\% statistical limits
on the field amplitude today $f_{R0}$ and in the bottom panel
the Compton wavelength in the background today $\lambda_{C0}$.  
The medium shaded region shows the
result for the default constraint, $|f_{R0}| < 1.3\times 10^{-4}$ at $n=1$,
using the modified spherical collapse parameters (lower edge of shaded
band in \reffig{nplot}).  
The dark region shows the most conservative constraints ($|f_{R0}|< 3\times 10^{-4}$), using the modified
collapse parameters and in addition assuming X-ray masses are underestimated
by 9\%.  Finally, the light region shows more aggressive constraints
($|f_{R0}| < 4\times 10^{-5}$),
using the unmodified collapse parameters (upper edge of shaded band in 
\reffig{nplot}).  Note that even this case is still somewhat conservative,
since for clusters at fixed mass, dynamical mass estimates such as X-ray 
masses will be enhanced by $\sim 20\%$ in the large-field limit of $f(R)$ gravity \cite{dynamicalmass}, due to the increased depth of the potential well.  
This increases the abundance at fixed $M_X$ in $f(R)$ considerably.

While the change in the fractional enhancement of the mass
function from $\Lambda$CDM with $n$ is relatively small, the impact on the model parameters can be large.  Specifically between the $n=1$ and $n=4$ models
the field amplitude limits change by over an order of magnitude and Compton wavelength constraints by a factor of several.

Nonetheless the cluster abundance measurements can already rule out a 
substantial portion of the cosmologically interesting regime for all cases, limiting the allowed range of
enhanced forces to $10-100$~Mpc.  Future large cluster samples
have the potential to push the limits down by an order of magnitude before 
chameleon effects cause a suppression of the enhancement.

\section{Discussion} \label{sec:discussion}

We have conducted $N$-body simulations to test the enhancement of the
cluster abundance in a variety of $f(R)$ models.   These models differ
in the redshift evolution of both the linear force enhancement and the non-linear
chameleon mechanism which suppresses such enhancements in the
deep gravitational potential wells of clusters of galaxies.
These results test the robustness of model independent techniques such
as spherical collapse for predicting the enhancement and constraining
modified gravity with cosmological data.

We find that for cluster mass halos, the spherical collapse predictions work
equally well for different models at least as long as the redshift evolution of the field is not
so steep as to invalidate the division between large field and small field regime
imposed at $|f_{R0}| \approx 10^{-5}$ for the background field amplitude at $z=0$.   In the large field regime
the background field amplitude is larger than the depth of the gravitational potential
wells of clusters and hence the chameleon effect is inoperative.  For
a scaling index $n \simlt 4$, a large field model retains this property for
$z \simlt 1$ when clusters form.  In this regime, the fractional enhancement of the
cluster abundance relative to $\Lambda$CDM is a relatively weak function of
$n$ that is determined by the evolution of the Compton wavelength or range of the
force in the background.  In the opposite small field regime, the enhancements
become suppressed above a limiting mass that depends mainly on the
field amplitude $M_{\rm cham} \propto |f_{R0}|^{3/2}$.

We use these results to extend the implications of the local cluster abundance
to the whole class of broken power law models.
 Most models in the literature can be mapped onto the parameter space of this
class. 
 Constraints on the field amplitude and Compton wavelength today are strongly model dependent due to the logarithmic dependence of the cluster abundance on their values in 
 any given model.   
 
 Results based on different model assumptions can be mapped onto each other
 by matching instead the linear theory rms fluctuation at the radius implied by the observed mass scale or even more directly by matching spherical collapse mass function predictions as we have shown for the local $X$-ray cluster abundance.

 \smallskip
 \noindent {\it Acknowledgments}:   SF and WH were supported by
 the Kavli Institute for Cosmological Physics (KICP) at the University
 of Chicago through grants NSF PHY-0114422 and NSF PHY-0551142 and an
 endowment from the Kavli Foundation and its founder Fred Kavli. SF was additionally supported by Eugene and Niesje Parker and Robert G. Sachs fellowships.   WH
 was additionally supported by U.S.~Dept.\ of Energy contract
 DE-FG02-90ER-40560 and the David and Lucile Packard Foundation. 
 FS was supported by the Gordon and Betty Moore Foundation at Caltech.  
  Computational resources for the cosmological simulations were provided by the KICP-Fermilab
computer cluster.

\bibliographystyle{arxiv_physrev}

\bibliography{frmodels}

\end{document}